\documentclass[fleqn,10pt]{wlscirep}
\usepackage[utf8]{inputenc}
\usepackage[T1]{fontenc}
\usepackage{braket}

\title{Ground state and collective excitations of a dipolar Bose-Einstein condensate in a bubble trap}

\author[1]{Pedro C. Diniz}
\author[1]{Eduardo A. B. Oliveira}
\author[2,$\dagger$]{Aristeu R. P. Lima}
\author[1,*]{Emanuel A. L. Henn}
\affil[1]{Institute of Physics of S\~ao Carlos, University of S\~ao Paulo,
PO Box 369, 13560-970, S\~ao Carlos, SP, Brazil}
\affil[2]{Universidade da Integração Internacional da Lusofonia Afro-Brasileira, Campus das Auroras, Acarape-Ceará, Brazil}

\affil[$\dagger$]{aristeu@unilab.edu.br}
\affil[*]{ehenn@ifsc.usp.br}

\begin{abstract}
We consider the ground state and the low-lying excitations of dipolar Bose-Einstein condensates in a bubble trap, i.e., a shell-shaped spherically symmetric confining potential. By means of an appropriate Gaussian ansatz, we determine the ground-state properties in the case where the particles interact by means of both the isotropic and short-range contact and the anisotropic and long-range dipole-dipole potential in the thin-shell limit. Moreover, with the ground state at hand, we employ the sum-rule approach to study the monopole, the two-, the three-dimensional quadrupole as well as the dipole modes. We find situations in which neither the virial nor Kohn's theorem can be applied. On top of that, we demonstrate the existence of anisotropic particle density profiles, which are absent in the case with repulsive contact interaction only. These significant deviations from what one would typically expect are then traced back to both the anisotropic nature of the dipolar interaction and the novel topology introduced by the bubble trap.
\end{abstract}
\begin{document}

\def\edd{\epsilon_{\rm dd}}          

\flushbottom
\maketitle
%
%
\thispagestyle{empty}

\section{Introduction}

The realization of Bose-Einstein condensation inaugurated a fertile and ever growing research field in physics. First obtained in dilute atomic gases \cite{anderson_mh_1995,davis_kb_1995}, Bose-Einstein condensates (BEC) have provided a series of remarkable breakthroughs. In a far from exhaustive list, one could include the observations of vortex-lattices \cite{J.R.Abo-Shaeer04202001}, the BCS-BEC crossover \cite{greiner_jin}, the Mott to superfluid quantum phase transition in an optical lattice \cite{mott}, the Bose-nova collapse \cite{bosenova}, and more recently of the supersolid state in dipolar gases \cite{modugno_supersolid,pfau_supersolid,ferlaino_supersolid}. In particular for the supersolid systems, beyond mean-field physics has been shown to play a crucial role so that studies including quantum fluctuations in dipolar BECs \cite{Lima_qf_1,Lima_qf_2} concerning the ground-state and excitations \cite{falk_ground_state}, the self-bound character of the droplet solutions \cite{blakie_self_bound} as well as vortices \cite{cidrim} have been carried out. 

Very commonly, breakthroughs are associated to the introduction and/or a higher level of control upon interaction terms \cite{Griesmaier2006a,Lahaye2007, spielman} or the control of the trapping potential landscape, either through the geometry of the system \cite{} or its dimensionality \cite{}. In special, quantum gases of Erbium\cite{} and Dysprosium \cite{} or Chromium close to a Feshbach resonance \cite{}, whose static and dynamic properties are dominated by dipole-dipole interactions (DDI), strongly profit from geometric and dimensional freedom in quantum gases systems: DDI of a polarized quantum gas is anisotropic, showing both attractive and repulsive characters, and long-range. \cite{muller2001}

 On a totally different perspective, the BEC physics might be on the verge of opening one further promising road. Indeed, in the absence of gravity the exploration of several phenomena is possible. The recent realization of a space-born BEC \cite{space} is a part of a large set of experiments planned for the microgravity conditions inside the Space Station. Moreover, a recent proposal to implement a realistic experimental framework for generating a BEC with shell geometry using radiofrequency dressing of magnetically-trapped samples has been made \cite{aveline}, opening further perspectives and reassuring the interest of the community.

In particular, BECs trapped in shell-shaped potentials would benefit in such microgravity environment: at Earth's surface, atoms in such trap just sag to the bottom of the shell \cite{foot_shell, nug_thesis, Perrin, white}. Indeed, the availability of such environment  triggered several theoretical efforts in order to unveil the collective modes and expansion dynamics in a bubble trap \cite{lannert}. Also, the hollowing transition, brought about by a suitable manipulation of the trap parameters, was shown to imprint its signature in the collective excitations of the system \cite{padavic}. On top of that, a recent systematic investigation of both the static and dynamic properties of shell-shaped BECs has been presented, which contains a comprehensive approach to the ground-state properties and low-lying excitations by means of both analytic and numerical results \cite{sun}. Recently, the fundamental aspects of Bose-Einstein condensation itself in the surface of a sphere had been investigated \cite{monica_shell} {\bf together with the possibility of cluster formation \cite{prestipino} and} the superfluid properties are studied in different regimes, including the Berezinski-Kosterlitz-Thouless phase transition \cite{salasnich}.

The current efforts aiming for a deeper understanding of shell-trapped BECs share an important feature: the atoms interact only \textit{via} the short range and isotropic contact interaction. The investigation of BECs displaying anisotropic dipole-dipole interactions, trapped in spherically symmetric thin shells is a natural extension of such a problem that presents unique characteristics: while the trapping is locally quasi-2D, the dipole-dipole interaction remains 3D and its anisotropic character breaks the spherical symmetry of the system. The ground state and stability parameters of such configuration have been investigated numerically\cite{adhikari} for a very specific set of trapping parameters in a more general context that focused on rings and vortices.

The present work is concerned with shell-shaped BECs featuring the long-range and anisotropic dipole-dipole interaction (DDI) in thin-shell limit (TSL) of a strong bubble trap without gravity. We choose to focus on this limit, as it highlights the particular effects brought about by the interplay between the bubble trap and the dipole-dipole interaction. We show that both the static and dynamical properties of the system are modified while we still recover results from previous publications without dipolar interactions. In the following, we investigate the ground-state configuration as well as the most important excitation modes.

\section{Results}
In this section, we present our approach to a dipolar Bose gas in a bubble trap in the thin-shell limit, where the width of the spherical shell is much smaller than the corresponding radius. In this regime, the most important features which are uniquely attached to the DDI can best be highlighted.

\subsection{Variational approach}

Consider a set of N bosonic dipoles aligned along the z-direction, possessing mass $M$ and trapped in a potential of the form
\begin{equation}
 U_{\textrm{B}}({\bf x}) = \frac{1}{2}M\omega_{0}^{2}(r-r_{0})^{2},
\end{equation}
which corresponds to a bubble potential \cite{zobay_1,zobay_2}, where the average radius $r_{0}$ and the oscillation frequency $\omega_{0}$ can be experimentally tuned \cite{sun}. Notice that one can define an oscillator length corresponding to the usual form $a_{\textrm{osc}} = \sqrt{{\hbar}/{M\omega_{0}}}$, with $\hbar$ being the reduced Planck constant.

The full interaction potential reads
\begin{equation}
V_{int}({\bf x}) = g \delta({\mathbf x}) + \frac{C_{\rm dd}}{4\pi|{\bf x}| ^{3}}\left[1-3\frac{z^{2}}{|{\bf x}|^{2}}\right]
\label{DDI}
\end{equation}
where $g={4\pi\hbar^{2}a_{\textrm{s}}}/{M}$ characterizes the strength of the usual short-range and isotropic contact interaction with s-wave scattering length $a_{\textrm{s}}$, while the second term stands for the long-range and anisotropic dipole-dipole interaction for dipoles polarized in the z-direction. Here $C_{\rm dd}$ is a constant related to the strength of the dipoles (either magnetic or electric). Moreover, we define $\edd = C_{\rm dd}/(3g)$ as the relative magnitude of the interaction.

Within this framework, the total Gross-Pitaevskii energy is given by
\begin{equation}
{ E}_{\textrm{GP}}[{\Psi_{}}] = E_{\textrm{kin}} + E_{\textrm{B}} + { E}_{int}
\label{energy_func}
\end{equation}
with the one-body part consisting of the kinetic
\begin{eqnarray}
E_{\textrm{kin}} & = & \frac{\hbar^{2}}{2M} \int{{\rm d}^{3}x}\,{\nabla\Psi^{*}}({\bf x})\cdot{\nabla\Psi^{}}({\bf x})
\label{ekin}
\end{eqnarray}
and the bubble trapping energy
\begin{eqnarray}
E_{\textrm{B}} & = & \int{{\rm d}^{3}x}|\Psi({\bf x})|^{2}U_{\rm B}({\bf x}).
\end{eqnarray}
The interaction energy, in turn, is given by
\begin{equation}
{ E}_{\textrm{int}} = \frac{1}{2}\int{{\rm d}^{3}x'}{{\rm d}^{3}x}|\Psi({\bf x})|^{2}{V_{\textrm{int}}({\bf x}-{\bf x'})}|\Psi({\bf x}')|^{2}.
\label{ddi_ener}
\end{equation}

\subsection{Spherical ansatz in the thin-shell limit}

In previous studies, where only the short-range and isotropic contact interaction was present, a spherically symmetric ansatz for the wave function was used \cite{padavic}. Due to the presence of the DDI, however, one looses the spherical symmetry. Therefore, we apply a normalized trial wave function which is capable of exhibiting possible corresponding changes in the cloud profile
\begin{equation}
\Psi_{\textrm{Trial}}({\bf x}) = \frac{\sqrt{N}}{\sqrt[4]{\pi}R_{0}\sqrt{R_{1}}}{\mathcal{F}}\left(\frac{r-R_{0}}{R_{1}}\right)\times h(\theta,\phi),
\label{ansatz_0}
\end{equation}
such that the radial part features a Gaussian distribution ${\mathcal{F}}(x) =  e^{-x^{2}/2}$ and the angular part is given in terms of spherical harmonics
\begin{eqnarray}
h(\theta,\phi) & = & \sum_{l,m}a_{l,m}Y_{l_{}}^{m}(\theta,\phi)
\end{eqnarray}
with normalization $\sum_{l,m}a_{l,m}^{}a_{l,m}^{*} =  1$.

 In a filled sphere, the usual effect of the DDI is to elongate the cloud along the polarization direction of the dipoles, as demonstrated previously in both bosonic and fermionic systems (see \cite{lahaye-review,baranov-review} and references therein). In a thin spherical shell, with the width much shorter than the radius, the distance between particles in different parts of the sphere renders this effect negligible and the DDI becomes mainly responsible for the rearrangement of the particles over the shell.

In what follows, we restrict ourselves to the thin-shell limit (TSL), in which most of the particles are at distance $R_{0}$ from the origin. Therefore, in the thin-shell limit, we apply the ansatz (\ref{ansatz_0}) and retain only the leading terms in $R_{1}/R_{0}$ in the total energy.  Under typical experimental conditions, this limit can be realized even in the Thomas-Fermi approximation \cite{lannert, sun}. The latter, however, is not assumed here.

\subsection{Ground-state configurations}

\begin{figure}[t]
\centering
\includegraphics{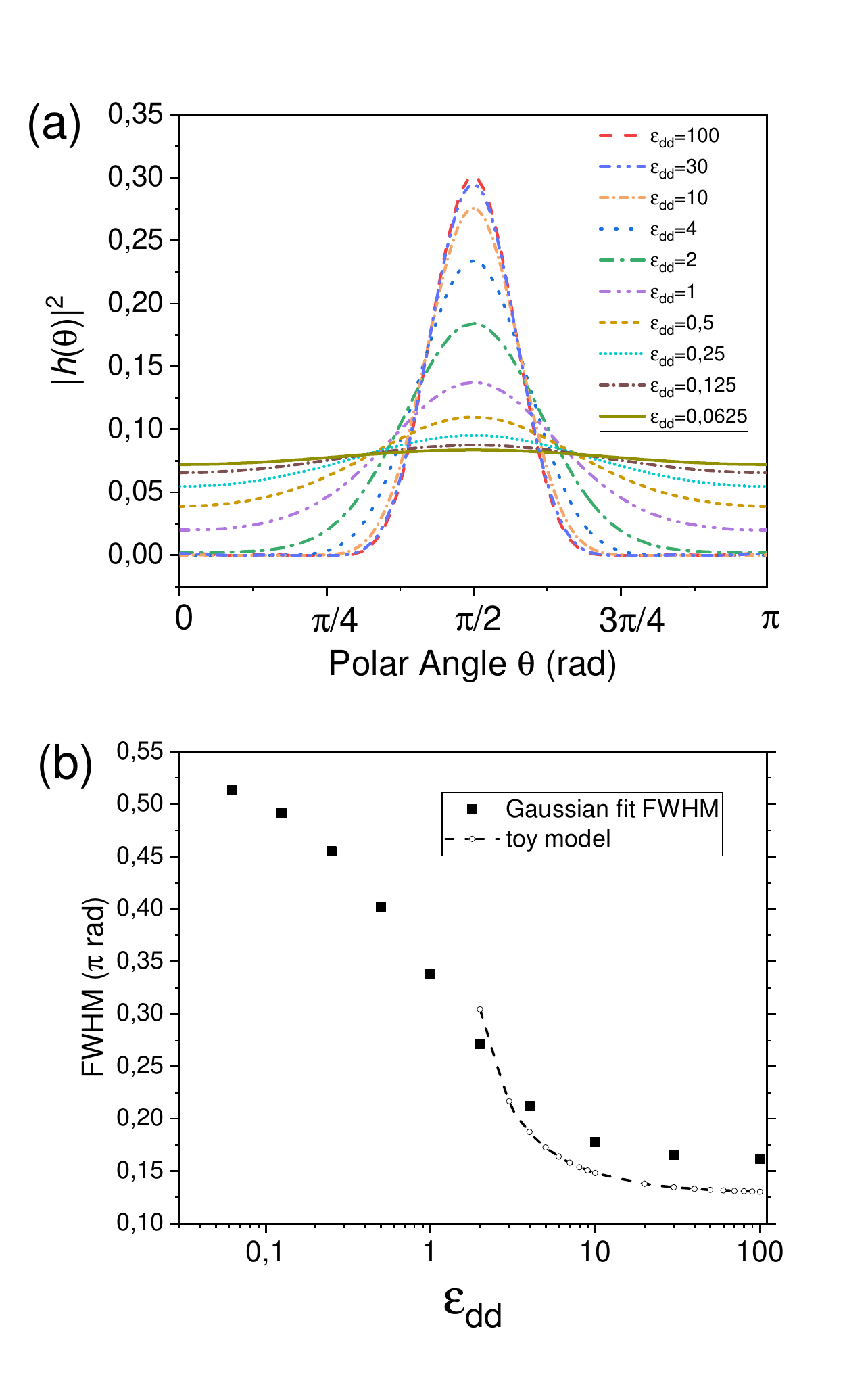}
\caption{(a) Polar distribution of the ground-state particle density $|h(\theta)|^{2}$ as a function of the polar angle $\theta$ for different values of $\edd$.
The larger the value of $\edd$, the more the particles tend to accumulate along the equator of the sphere. (b) Full width at half maximum of gaussian fits to the ground-state distributions of (a) as a function of $\edd$ (squares) showing the tendency of the ground state to saturate at a minimum width. Dashed line with open dots are the same quantities obtained by our toy model (see text). }
\label{fig:ground}
\end{figure}

Performing a numerical minimization of the energy (\ref{energy_func}) with respect to the coefficients of expansion (\ref{ansatz_0}), we obtain the ground-state configuration of the system. In the absence of the DDI, the particle density reflects the spherical symmetry of the trap. For non-vanishing $\edd$, however, the orientation axis of the dipoles constitutes a preferred direction so that the particles rearrange correspondingly. While the spherical symmetry is broken, the azimuthal symmetry around the polarization axis remains.

For definiteness, we choose experimentally realistic values for the parameters which represent a feasible finite thin shell. {\bf We consider $10^{4}$ particles and then constrain the radial coordinate to be $r=R_0$ and adopt $R_{0}=20a_{\textrm{osc}}$ and $R_{1}=a_{\textrm{osc}}$, so that one has $R_{1}=R_{0}/20$. Also, $\omega_0=2\pi \times 200Hz$. In a harmonically trapped, even in the Thomas-Fermi regime, the condensate radius is just a few times the oscillator length. In the case of a bubble trap, the Thomas-Fermi shell width is actually much smaller \cite{lannert}. Therefore, our choice for the shell parameters is, indeed, reasonable. Morover, these parameters are in the range of those used in the TSL of Ref.\cite{sun} in order} to allow for a direct comparison, where possible, in the vanishing dipole-dipole interaction limit. Therefore, the variational parameters of interest in the thin-shell limit are contained in the angular part $|h(\theta,\phi)|^{2}$, which we proceed to optimize numerically by minimizing the total energy. More details are described in the methods section below. As expected, there is no dependence on the azimuthal $\phi$ angle.

In Fig. \ref{fig:ground}(a), we show angular distribution of ground-state density $|h(\theta)|^{2}$ over the sphere as a function of the polar angle for several values of $\epsilon_{dd}$. We see that, for increasing values of the dipolar strength $\edd$, the density becomes larger at the equator and it eventually vanishes at the poles. In Fig.\ref{fig:ground}(b) we quantify this effect by doing a simple gaussian fit to the angular distributions and plotting full width at half minimum (FWHM) of the angular distribution as a function of $\epsilon_{dd}$. While for small $\edd = 0.0625$, the width amounts to $\approx 0.5\pi$ rad, it saturates to a minimum value around $0.17\pi$ rad, as $\edd$ is increased to a very large value $\edd = 100$.

We interpret this result in terms of the pictorial representation of the DDI, according to which dipoles aligned along a given direction tend do repel each other, if they are oriented side by side, while an attraction takes place between them in a head-to-tail orientation. In the bubble trap, dipoles along the equator experience attraction from other dipoles located above and below them along the meridian lines while they are repelled by the ones along the equator. Dipoles located at the poles, on the contrary, only experience repulsion from the surrounding particles. Therefore, a configuration in which more particles are on the equator leads to a lower total energy.

To support this interpretation and gain some insight on the problem, we have developed a toy model focusing on the particle density around the equator. We consider dipolar particles confined in a thin rectangular plate, such that the direction with the shortest length (y) is perpendicular to the polarization direction (z). We then assume Gaussian density distributions in both z and y directions with corresponding widths $\sigma$ and $\beta$, respectively. The density along the third direction (x) is taken to be homogeneous inside the plate, for simplicity, and vanishing outside. Moreover, the length in the x-direction is taken to be finite at first, ranging from $x=-L$ to $x=L$. Later on, we take the limit $L\rightarrow \infty$ to mimic periodic boundary conditions. If one would roll such a thin plate around the z-axis to match the ends on x-direction, that would resemble the density distribution in a bubble trap in the TSL for large $\epsilon_{dd}$, as the BEC occupies a narrow, quasi-flat, region around the equator. In this case, the density in the y- and z-directions should satisfy $\beta \propto R_1$ and $\sigma \propto$ FWHM, respectively.

In this configuration, the interaction is the most important energy contribution, as kinetic and trapping energies are nearly frozen out. Therefore, we calculate the contact and dipolar interaction energies and obtain

\begin{equation}
U_{int}=\frac{g\lambda^2}{16\pi}\left(\frac{1}{\sigma\beta}+\epsilon_{dd}\frac{2\beta-\sigma}{\sigma\beta(\sigma+\beta)}\right)    
\end{equation}
where $\lambda$ is a constant obtained from the normalization of the density distribution to a given number of atoms N.

Minimizing $U_{int}$ with respect to $\sigma$ leads to a relation between the Gaussian length in the z-direction $\sigma_{min}$ and the plate width $\beta$

\begin{equation}
    \sigma_{min}=\beta\left(\frac{1+2\epsilon_{dd}+\sqrt{3\epsilon_{dd}(1+2\epsilon_{dd})}}{\epsilon_{dd}-1}\right),
\end{equation}
which could throw light upon the particle concentration on the equator. We plot this expression from $\epsilon_{dd}=2$ to $\epsilon_{dd}=100$ as a dashed line with open dots in Fig.\ref{fig:ground}(b) also displaying an asymptotic behavior at large $\epsilon_{dd}$. Indeed, what we find is a good overall agreement over nearly two orders of magnitude, despite neglecting the one-body energy contributions.

We remark that this simple sheet-like toy model loses validity as we approach $\epsilon_{dd}=1$ from above, since the density distribution widens and starts to probe the curvature of the bubble, but the good quantitative agreement indicates that the interaction energy is responsible by this compression of the cloud towards equator in contrast to the filled trap, which elongates itself.

\subsection{Low-lying excitations}
Now that we have obtained the ground state of a dipolar BEC in a thin shell, we are in position to investigate the low-lying excitations of the system. We do so by means of the sum-rule approach, which has been applied successfully to both bosonic \cite{stringari_s_1996} and fermionic \cite{vichi_coll} gases in a harmonic trap. In this approach, an upper limit for the excitation energy of a given operator $F$, written in first quantized form, can be estimated through the ratio
\begin{equation}
\hbar \omega^{upper} = \sqrt{\frac{m_3}{m_1}}, \label{upper}
\end{equation}
where $m_i \equiv \sum_n    |\braket{0|F|n}|^2(\hbar \omega_{n0})^i $ is the i-th moment of the operator $F$. The convenience of the method lies in the fact that these moments can be put in the form
\begin{align}
m_1 &= \tfrac{1}{2}\braket{0\bigl| [F^\dagger,[H,F]] \bigl|0}, \\[1mm]
m_3 &= \tfrac{1}{2}\braket{0\bigl| \bigl[[F^\dagger ,H],[H,[H,F]] \bigl] \bigl|0},
\end{align}
where the expectation values are to be calculated with respect to the ground state. Moreover, the hamiltonian $H=\sum_{i}\left[p_{i}^{2}/2m + U_{\rm B}({\bf r}_{i})\right]+\sum_{i<j}{V_{\textrm{int}}({\bf r}_{i}-{\bf r}_{j})}$ is also written in first quantized form.

\subsubsection{Two-dimensional quadrupole mode}

The two-dimensional quadrupole mode corresponds to a vibration, such that the oscillations are out of phase in the xy-plane while the z direction remains frozen. It is excited by the operator $F_{|m|=2} = \sum_i (x_i^2-y_i^2)$ and we obtain that, in general, its frequency is given by
\begin{equation}
\omega_{|m|=2} = \sqrt{\frac{{ 2 E_ {kin \perp}+2NM\omega_{0}^{2}\braket{0\bigl| r_{\perp}^{2} \left(1-\frac{r_{0}}{r}\right)\bigl|0}}}{NM\braket{0\bigl| x^2 \bigl|0}} },
\label{rd_quad_freq}
\end{equation}
where $E_ {kin \perp}$ and $r_{\perp}^{2}=x^{2}+y^{2}$ represent the kinetic energy and the square radius in the xy-plane. Here, we have used equation (\ref{upper}) for the frequency, which therefore consists in an upper bound. We remark that such results, however, are usually indistinguishable from the ones given by other methods.

Since we are working in the thin-shell limit, the ground state is concentrated in the region $r\approx r_{0}$. Therefore, we obtain
\begin{equation}
\omega_{|m|=2} \cong  \sqrt{\frac{ 2 E_{kin \perp}}{NM \braket{0\bigl|x^2\bigl|0}}}.
\label{rd_quad_freq_ths}
\end{equation}

Due to the usual precision with which excitation frequencies are measured (a few Hz), comparison between (\ref{rd_quad_freq}) and (\ref{rd_quad_freq_ths}) provides an useful experimental tool to determine the achievement of the TSL. The prospects for detecting the influence of the DDI in this regime are, however, not very promising, as the difference appears only in the first decimal place as one ranges from a very strongly dipolar system ($\edd^{-1}\rightarrow 0$) to a virtually non-dipolar one (large $\edd^{-1}$), as shown in Fig. \ref{fig:multiplefreq} (red circles, dashed line). Notice that in Fig. \ref{fig:multiplefreq} we plot excitation frequencies as a function of $\edd^{-1}$ so the horizontal axis is directly proportional to the s-wave scattering length $a_s$ which is the experimentally accessible quantity to manipulate while maintaining the possibility to scale our results to any dipolar system.  We remark that, as $\edd$ tends to zero, our result approaches the non-dipolar excitation frequency\cite{sun} obtained \textit{via} hydrodynamical equations very accurately and that such very low-frequency modes are characteristic of the TSL regime and non-existent in filled traps.

\subsubsection{Monopole and three-dimensional quadrupole modes}

\begin{figure}[ht]
\centering
\includegraphics[width=\linewidth]{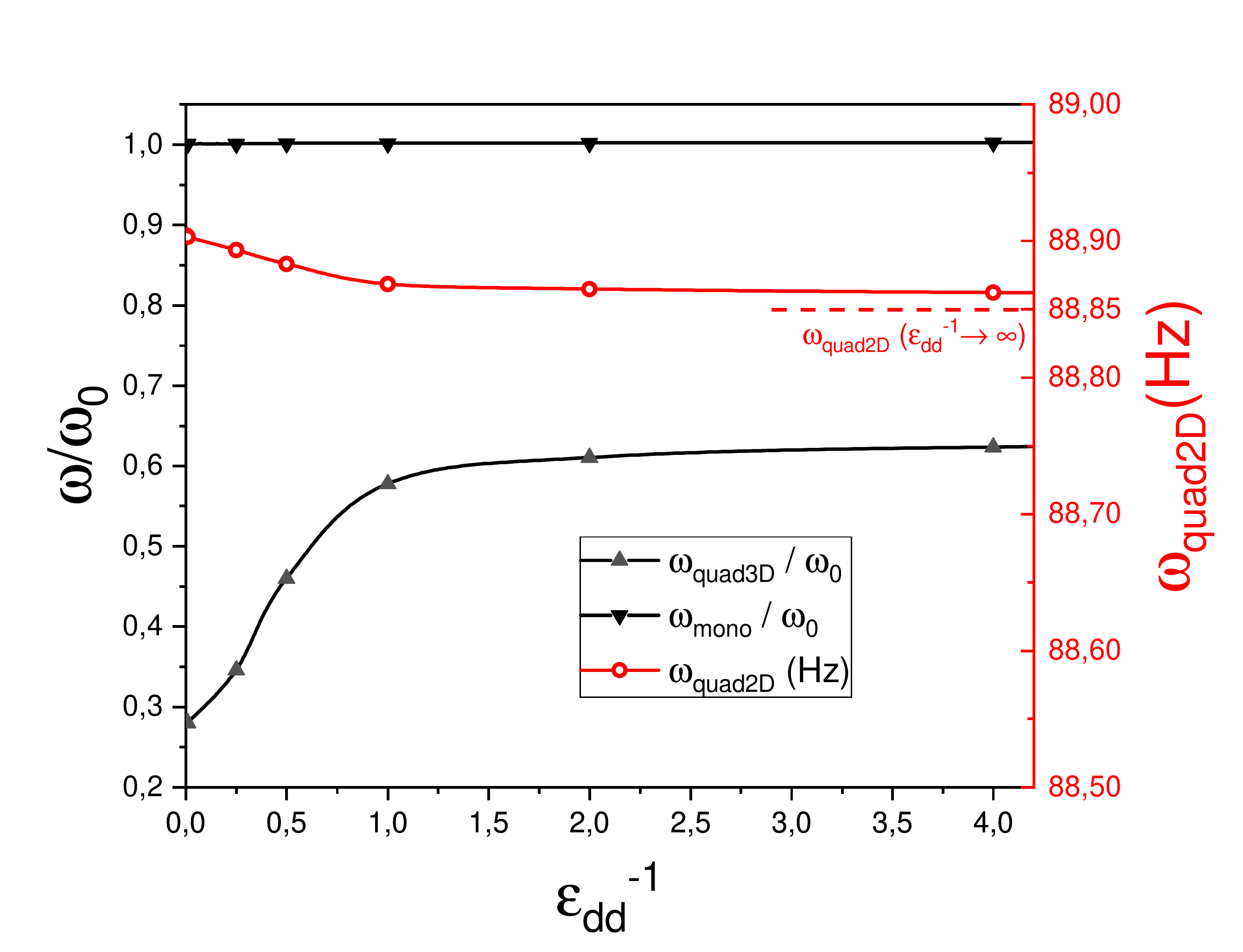}
\caption{Monopole (downward black triangles) and three-dimensional quadrupole (upward gray triangles) excitation frequencies in units of $\omega_{0}$ and two-dimensional quadrupole (red circle) in Hz, as functions of $\edd^{-1}$ for $\omega_0=2\pi \times 200$Hz and $\frac{R_0}{R_1}=20$. The curves serve as guides to the eye. For the two-dimensional quadrupole mode, we also indicate the non-dipolar frequency {\bf, calculated from the result of Ref. \cite{sun},} as a horizontal dashed line.}
\label{fig:multiplefreq}
\end{figure}

Let us now present the low-lying excitation frequencies for the monopole and three-dimensional quadrupole modes. The former is characterized by in-phase expansion and compression of the whole system, while the latter features out-of-phase oscillations in the radial and z-directions. In the absence of spherical symmetry, which is removed by the DDI, these modes are coupled. We follow a previous study \cite{kimura} and overcome this difficulty by using the operator $F= \sum_i\left( r_{\perp,i}^2 - \alpha z_{i}^2 \right)$, where the sum extends over all the particles. Then, the monopole (three-dimensional quadrupole) frequency is obtained by maximizing (minimizing) the upper limit (\ref{upper}) with respect to $\alpha$. The formulas obtained for the frequencies in this manner are not enlightening and we omit them while focusing on the graphical result exhibiting their dependence on the relative interaction strength $\edd$.

In Fig. \ref{fig:multiplefreq}, we show the ratio between the frequencies of the monopole and three-dimensional quadrupole modes and the trap characteristic frequency $\omega_0$ as a function of the dipolar interaction strength $\edd$. Notice that the monopole frequency remains unaltered for all practical purposes ($\Delta(\frac{\omega_{mon}}{\omega_0})\approx0.1\%$ over the whole range shown) although again the non-dipolar limit matches very well the one obtained in Ref.\cite{sun} through hydrodynamic equations ($\omega_{mon}\approx1.002\omega_0$). This is remarkably different from what happens in both fermionic and bosonic dipolar gases in harmonic traps. For a dipolar BEC in a harmonic trap, the monopole frequency is always larger for a dipolar gas than for a non-dipolar one \cite{odell_dhj_2004}, while dipolar Fermi gases in the hydrodynamic regime display similar behaviour \cite{lima_fermi_1,lima_fermi_2}.

The three-dimensional quadrupole frequency, on the other hand, displays, in the non-dipolar limit, a frequency much smaller than trap frequency, in contrast with the filled trap and also exhibits a substantial variation as $\edd$ increases ($\edd^{-1}\rightarrow 0$), marking a clear signal of the interaction upon the low-lying excitations in the bubble trap. For this reason, we remark that this mode is the most promising one with respect to the detection of the DDI in BECs in bubble traps. Notice that, for this mode, we do not have hydrodynamic calculations to compare with.



\subsubsection{Dipole mode}

\begin{figure}[t]
\centering
\includegraphics[width=\linewidth]{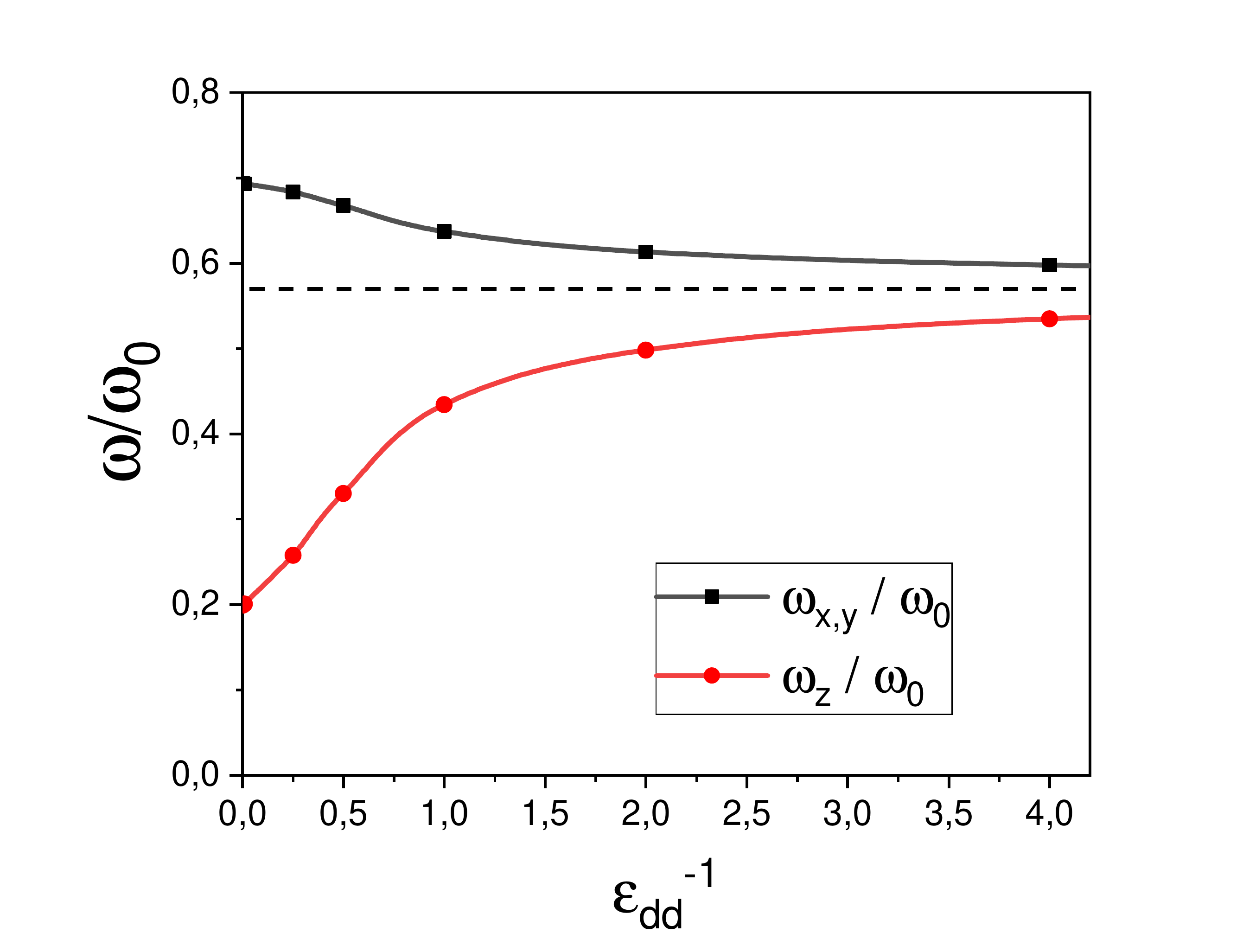}
\caption{Dipole mode excitation frequencies in units of $\omega_{0}$ as a function of $\edd^{-1}$.}
\label{fig:dipole}
\end{figure}

Let us now discuss the center-of-mass (COM) motion, excited by the operators $ F_x =  \sum_i x_i $ and $F_z =  \sum_i z_i $, whenever the motion is to take place in the x or z directions, respectively. In a harmonic trap, irrespective of the presence and nature of the interactions, the COM oscillates with the same frequency as the trapping potential, as demanded by Kohn's theorem. In a bubble trap, however, this is not the case. Using the sum-rule approach, we obtain
\begin{eqnarray}
\omega_{x}^{\rm SR}  & = & \omega_{0} \left( 1 - \frac{r_{0}}{N} \braket{0\bigl|\frac{1}{r}\bigl|0} + \frac{r_{0}}{N}\braket{0\bigl|\frac{x^{2}}{r^{3}}\bigl|0} \right)^{1/2},\quad \omega_{z}^{\rm SR} = \omega_{0} \left( 1 - \frac{r_{0}}{N} \braket{0\bigl|\frac{1}{r}\bigl|0} + \frac{r_{0}}{N}\braket{0\bigl|\frac{z^{2}}{r^{3}}\bigl|0} \right)^{1/2},
\label{dipole_sum_rule}
\end{eqnarray}
for the oscillation frequencies of the COM motion in the x and z directions, respectively. Notice that the direction in which the oscillations occur influences the frequency both explicitly, by means of the last term in the square root, and implicitly, through the expectation value in the ground state. The dipole frequencies in units of $\omega_0$ are shown in Fig. \ref{fig:dipole}. To the non-dipolar limit, all three frequencies are the same and equal to $\frac{\omega_i}{\omega_0}=\frac{\sqrt{3}}{3}\approx0.58$, since the ground state is isotropic. As the dipolar character increases ($\edd^{-1}\rightarrow 0$), $\omega_{x,y}$ increases while $\omega_z$ decreases. The softening of the axial COM motion as the atoms move away from the poles towards the equator of the bubble can be understood as the atoms probing an increasingly ``flat'' potential with lower effective trapping frequency along the polarization axis.

It is worth noting that in the TSL and for BECs with contact interaction only, expressions (\ref{dipole_sum_rule}) lead to a non-vanishing excitation frequency. This is in contrast to what is found for the dipole mode in the literature \cite{sun}, where the dipole oscillation frequency vanishes in the TSL. In order to understand this result better, we have investigated this mode also by means of a linearization of the density oscillations around the Thomas-Fermi density within the hydrodynamic approach \cite{pita_string}. In this configuration, an analytic solution can be obtained for both frequencies which are identical 
\begin{eqnarray}
\omega_{x,z}^{\rm HD}  & = & \omega_{0} \left( 1 - \frac{r_{0}}{N} \braket{0\bigl|\frac{1}{r}\bigl|0} \right)^{1/2},
\label{dipole_hd}
\end{eqnarray}
and differ from the sum-rule solutions by the additive term inside the square root in (\ref{dipole_sum_rule}). This term, on one hand, shows that the sum-rule solution gives a finite excitation frequency, even in the TSL, and, on the other hand, warrants that this solution is larger than the hydrodynamic one, as expected.

A word of caution is in order here, as the dipole mode is significantly modified by the presence of the DDI in a bubble trapped system. This feature is exclusively due to the shape of the trap, while the role of the DDI is seen in the anisotropy of the modification. Indeed, the frequency of the corresponding mode in a non-dipolar BEC has been found to change all the way from the trap frequency to zero, as the system is moved from a filled sphere to the thin-shell limit \cite{sun}. {\bf In addition, other situations have been found, in which Kohn's theorem cannot be applied. For instance, in photonic BECs \cite{enrico} and also in BECs with time-dependent scattering lengths \cite{edmir}.}

\section{Dicussion}

Bose-Einstein condensates in spherical bubble traps represent a recent major experimental achievement and have led to important theoretical developments in the context of the short-range and isotropic contact interaction. We have expanded the understanding of ultracold quantum gases by investigating the influence of the long-range and anisotropic dipole-dipole interaction in the limit of a thin shell, with the dipoles along the z-direction. By means of a Gaussian ansatz for the radial part of the wave function and a spherical harmonics expansion for the angular part, we were able to obtain analytic expressions for the total energy, which were then minimized with respect to variational parameters. Concerning the ground state, we have found that the equilibrium configuration displays azimuthal symmetry and the particles tend to accumulate along the equator of the sphere, an effect which can be best demonstrated in the absence of gravity. This reflects the fact that the DDI only distinguishes one direction, namely that of the dipoles. This is a key feature of the thin-shell limit, as in the case of a filled shell, particles tend to assume head-to-tail orientations, thereby stretching the cloud along the dipolar directions.  We have confirmed this tendency by means of a sheet-like model, mimicking the vicinity of the equator in the situation of a spherical shell with an infinite ratio between its radius and its width. The low-lying excitations were investigated with the help of the sum rule approach \cite{stringari_s_1996,vichi_coll,kimura}. Significant deviations with respect to the non-dipolar cases have been demonstrated, providind important evidence for the experimental detection of both excitation properties of the system and the onset of the TSL. As a result, the first demonstration of dipolar effects in bubble trapped Bose gases, as carried out here, can serve as a guide to future theoretical as well as experimental investigations.

\section{Methods}

Applying ansatz (\ref{ansatz_0}) and neglecting terms of order $R_{1}^{2}/R_{0}^{2}$, we obtain the following expressions for the trapping and kinetic energies
\begin{eqnarray}
E_{\textrm{B}} & = & \frac{N M\omega_{0}^{2} }{2}\left(R_{0}-r_{0}\right)^{2},\quad E_{\textrm{Kin}} = \frac{N\hbar^{2}}{2M}\left[\frac{1}{2}\frac{1}{R_{1}^{2}} + \sum_{l,m}a^{}_{l,m}a_{l,m}^{*}\frac{l(l+1)}{R_{0}^{2}}\right],
\end{eqnarray}
respectively, so that the former is minimized by requiring that $R_{0}=r_{0}$. For this reason, for a sufficiently strong trap, particles tend to accumulate at a fixed distance $R_0$ of the center, thereby causing a hole in the cloud. This changes completely the properties of the system and has important consequences. Notice that, for vanishing $l$, the radius of the sphere plays no role and all the kinetic energy is stored in the shell width. Moreover, for non-vanishing $l$, the second term in the kinetic energy agrees with the energy of a particle in a sphere of radius $R_{0}$ \cite{salasnich}.

The short-range interaction energy reads
\begin{eqnarray}
E_{\delta} & = & \frac{g{N}^{2}}{2\sqrt{2\pi} R_{0}^{2}R_{1}}\sum_{l_{1},m_{1},l_{2},m_{2},l_{3},m_{3},l_{4},m_{4}} a_{l_{1},m_{1}}a_{l_{2},m_{2}} a^{*}_{l_{3},m_{3}}a^{*}_{l_{4},m_{4}} I_4 (l_{1},m_{1},l_{2},m_{2},l_{3},m_{3},l_{4},m_{4})\nonumber\\
\end{eqnarray}
with the auxiliary coefficient $I_4$ being discussed in the Appendix. Here, we remark that these coefficients being explicitly positive for $m=0$ leads to $l=0$ being a preferred state.

The DDI energy is given by
\begin{eqnarray}
E_{\rm dd} 
& = & \frac{8 N^{2}}{\pi\sqrt{5}R_{0}^{2}R_{1}}  \frac{C_{\rm dd}}{3}  \sum_{l_{1},m_{1},l_{2},m_{2},l_{3},m_{3},l_{4},m_{4},l_{5},m_{5},l_{6}} a_{l_{1},m_{1}}a_{l_{2},m_{2}}^{*}a_{l_{4},m_{4}}a_{l_{5},m_{5}}^{*}  I_{DD}   \end{eqnarray}
with
\begin{eqnarray}
I_{DD} & = & I_3(l_{1},m_{1},l_{2},m_{2},l_{3},m_{3})    I_3(l_{4},m_{4},l_{5},m_{5},l_{6},m_{3}) (-1)^{m_{3}} I_3(l_{3},m_{3},2,0,l_{6},m_{3}) \nonumber\\
&&\frac{\pi}{2\sqrt{2}}\left\{\delta_{l_{6},l_{3}} + \left[1-(2l_{3}+3)\frac{R_{0}}{R_{1}}\sqrt{\frac{\pi}{2}} \right]\delta_{l_{6},l_{3}+2} +\left[1-(2l_{3}-1)\frac{R_{0}}{R_{1}} \sqrt{\frac{\pi}{2}} \right]\delta_{l_{6},l_{3}-2}
\right\} .
\end{eqnarray}
Notice that the DDI has angular momentum-conserving contributions, which resembles the contact ones and have no influence from $\frac{R_{0}}{R_{1}}$-terms. In addition, it also contains contributions which connect states with different angular momentum, which is an exclusive feature of anisotropic interactions.

{\bf We implement the TSL numerically for $10^{4}$ particles by choosing the values $\omega_{0} = 2\pi \times 200$Hz for the bubble trap frequency, $R_{0} = r_{0}= 20 a_{\textrm{osc}}$ for the trap radius, and $R_{1} = R_{0}/20$ and evaluate all our ground-state expectation values for this set of parameters. On top of that, we fix the dipolar strength $C_{\rm dd}$ and vary the s-wave scattering length so as to obtain a variation in the relative magnitude $\edd = C_{\rm dd}/(3g)$. This is justified, since actual experiments are carried out in this way, with the help of Feshbach resonances.}

\appendix

\section{\label{Ifour} Matrix elements of the interaction terms}

Let us briefly state the matrix elements which were used t obtain both the dipolar and contact interactions.

The coefficient $I_{4}$ can be evaluated analytically by means of standard techniques and we obtain
\begin{eqnarray}
I_4(l_{1},m_{1},l_{2},m_{2},l_{3},m_{3},l_{4},m_{4}) & = & \sum_{l_{}}\sqrt{\frac{(2l_{1}+1)(2l_{2}+1)}{4\pi}}
\sqrt{\frac{(2l_{3}+1)(2l_{4}+1)}{4\pi}} (2l_{}+1)
\nonumber\\
&&
\left(
\begin{array}{ccc} 
l_{1} & l_{2} & l_{}\\
0 & 0 & 0
\end{array}
\right)
\left(
\begin{array}{ccc} 
l_{1} & l_{2} & l_{}\\
m_{1} & m_{2} & -(m_{1}+m_{2})
\end{array}
\right)\nonumber\\
&&\times
\left(
\begin{array}{ccc} 
l_{3} & l_{4} & l_{}\\
0 & 0 & 0
\end{array}
\right)
\left(
\begin{array}{ccc} 
l_{3} & l_{4} & l_{}\\
m_{3} & m_{4} & -(m_{1}+m_{2})
\end{array}
\right)\delta_{m_{1}+m_{2},m_{3}+m_{4}},
\end{eqnarray}
where $\left(
\begin{array}{ccc} 
l_{1} & l_{2} & l_{}\\
m_{1} & m_{2} & m
\end{array}
\right)$ denotes the the Wigner 3-j symbol.

Proceeding in an analogous manner with respect to $I_{3}$ leads to
\begin{eqnarray}
 I_3(l_{1},m_{1},l_{2},m_{2},l_{3},m_{3})
& = & (-1)^{m_{1}} \sqrt{\frac{(2l_{1}+1)(2l_{2}+1)(2l_{3}+1)}{4\pi}}
\left(
\begin{array}{ccc} 
l_{1} & l_{2} & l_{3} \\
0 & 0 & 0
\end{array}
\right)
\left(
\begin{array}{ccc} 
l_{1} & l_{2} & l_{3} \\
-m_{1} & m_{2} & m_{3} 
\end{array}
\right).
\end{eqnarray}

\section*{Acknowledgements}

We thank F. E. A. dos Santos, A. Pelster, and S. Stringari for fruitful discussions.
This work was supported by the São Paulo Research Foundation (FAPESP) under the grants 2015/20475-9 and 2013/07276-1. P. C. D. acknowledges support of CNPq scholarship.  E. A. B. acknowledges support through FAPESP scholarship 2018/16369-7. A.~R.~P.~Lima acknowledges financial support from the Brazilian Funda\c{c}\~ao Cearense de Apoio ao Desenvolvimento Cient\'ifico e Tecnol\'ogico (Grant No.~BP3-0139-00281.01.00/18).

\section*{Author contributions statement}

E.A.L.H. and A.R.P.L. conceived the idea and supervised the work. E.A.L.H. developed the toy model. A.R.P.L. developed the fundamentals of the theory and overall theoretical guidance. P.C.D. carried out numerical calculations, E.A.B.O. developed the analytical expressions for the low-lying excitations. All authors analysed and interpreted the results. All authors wrote and reviewed the manuscript.

\end{document}